\documentclass[conference]{IEEEtran}
\usepackage{amsmath}
\usepackage{amsfonts}
\usepackage{amssymb}
\usepackage{epsfig}
\usepackage{epstopdf}
\usepackage{graphicx}
\usepackage{subfig}
\usepackage{tabularx}
\usepackage{url}
\usepackage{multirow}
\usepackage[section]{placeins}

\begin{document}
\title{Non-data-aided SNR Estimation for QPSK Modulation in AWGN Channel}

\author{
	\IEEEauthorblockN{Tara Salman \IEEEauthorrefmark{3}, Ahmed Badawy\IEEEauthorrefmark{1}\IEEEauthorrefmark{2}, Tarek M. Elfouly \IEEEauthorrefmark{3}, Tamer Khattab \IEEEauthorrefmark{2}, and Amr Mohamed \IEEEauthorrefmark{3} }

	\IEEEauthorblockA{\IEEEauthorrefmark{1}Politecnico di Torino, DET - iXem Lab. (ahmed.badawy@polito.it)}
	\IEEEauthorblockA{\IEEEauthorrefmark{2} Qatar University, Electrical Engineering Dept. }
	\IEEEauthorblockA{\IEEEauthorrefmark{3} Qatar University, Computer Science and Engineering Dept. }
    \IEEEauthorblockA{Qatar University\\
	Doha, Qatar 2713\\
    {tara.s, badawy, tarekfouly, tkhattab, amrm}@qu.edu.qa}
}
\maketitle
\begin{abstract}
Signal-to-noise ratio $($SNR$)$ estimation is an important parameter that is required in any receiver or communication systems. It can be computed either by a pilot signal data-aided approach in which the transmitted signal would be known to the receiver, or without any knowledge of the transmitted signal, which is a non-data-aided $($NDA$)$ estimation approach. In this paper, a NDA SNR estimation algorithm for QPSK signal is proposed. The proposed algorithm modifies the existing Signal-to-Variation Ratio $($SVR$)$ SNR estimation algorithm in the aim to reduce its bias and mean square error in case of negative SNR values at low number of samples of it. We first present the existing SVR algorithm and then show the mathematical derivation of the new NDA algorithm. In addition, we compare our algorithm to two baselines estimation methods, namely the M2M4 and SVR algorithms, using different test cases. Those test cases include low SNR values, extremely high SNR values and low number of samples. Results showed that our algorithm had a better performance compared to second and fourth moment estimation (M2M4) and original SVR algorithms in terms of normalized mean square error $($NMSE$)$ and bias estimation while keeping almost the same complexity as the original algorithms. 
\end{abstract}
\section{Introduction}
\label{S:Introduction}
A good SNR estimation is critical in many digital communication systems as it is a key parameter in many receiver application such as decoding~\cite{summers1998snr},power control in multiple-access systems and channel assignment. Hence, various algorithms were proposed to compute an accurate estimation of this parameter. In general, these algorithms can be divided into two main categories: data-aided $($DA$)$ and non-data-aided $($NDA$)$ estimation. A DA estimator, such as Maximum Likelihood $($ML$)$~\cite{1} and Squared Signal-to-Noise Variance $($SNV$)$~\cite{2}, would require the transmitted data to be perfectly known to the receiver, or at least the first few samples. As for the NDA estimators, such as second and fourth moment $($M2M4$)$ and Signal-to-Variation Ratio $($SVR$)$, they assume the transmitted signal to be unknown to the receiver~\cite{xu2005novel}. Even though the DA estimation would give a better and more accurate estimation~\cite{7}, the advantage of the NDA is that it does not need the transmitted data to be previously known at the receiver, hence it is more bandwidth efficient than the data-aided~\cite{wiesel2002non}. 

One common problem that was found experimentally in traditional NDA estimation algorithms is the high NMSE when SNR values are negative, or the noise power is higher than the signal power. Some researchers have tried to address this problem by modifying the existing well-established estimation methods. For example, in~\cite{3}, the authors proposed a modification on M2M4 algorithm by introducing another fourth moment representation that considers the in-phase and quadrature component difference instead of their summation. By that, the algorithm can reduce the bias and NMSE of the estimated SNR compared to the original M2M4 method. Results showed a better performance of the modified algorithm compared to the original M2M4 in the case of limited number of samples at low SNR values. 

Another problem in any SNR estimation is the complexity of the algorithm which is needed to be minimized in order to be practically implementable on any energy efficient hardware, such as FPGA. One work, presented in~\cite{4}, addresses the complexity of M2M4 and tries to reduce it by evaluating the noise power with the absolute values of the in-phase and quadrature components of the received signal instead of using the fourth moment. The complexity was analyzed and proven to be deducted by  at least 50\% while the performance was better than other NDA SNR estimations with the same complexity. Another work, presented in~\cite{5}, proposes a simplified structure of SNR estimation that tries to achieve a good performance with less computational complexity. The main idea of this work is to reformulate the noise in another way and consider the difference between the absolute values of the in-phase and quadrature components. Results showed that the proposed algorithm has a moderate performance that is slightly worse than M2M4 and better than the best method presented in~\cite{6} over the entire SNR values of interest.

In this paper, it was noted that most of the existing methods in NDA SNR estimation have a high NMSE when SNR values are low at low number of samples. Besides, in~\cite{7}, it was shown that SVR estimation performs worse than others in all SNR values of interest. Hence, this paper proposes an algorithm that modifies the SVR algorithm and tries to reduce the bias and NMSE of the estimation especially with limited number of samples at low SNR values. The algorithm was applied to a QPSK modulated signal in AWGN channel and can be applied to M-PSK modulated signal with some parametric changes. 

The rest of the paper will be organized as follows: section~\ref{sec:Original SVR Method} will present the basic SNR estimation concept with the mathematical derivation of the original SVR estimation algorithm, section~\ref{sec:proposed algorithm} will highlight the new algorithm mathematical derivation. Results of the new algorithm will be compared to original methods with different test cases in section~\ref{sec:results}. Finally, conclusion and future work will be presented in section~\ref{sec:Conclusion}. 
\section{Original SNR Estimation Algorithm}
\label{sec:Original SVR Method}
\subsection{System Model}
Consider QPSK received signal that can be expressed as: 
\begin{equation}
y=s+n
\end{equation}
where $s$ is the original signal that was sent by the transmitter and $n$ is added noise. 
Considering the in-phase and quadratic component of both signal and noise 
\begin{equation}
y=s_I+s_Q+n_I+n_Q	
\end{equation}
where $s_I,s_Q ,n_I,n_Q $ are the in-phase signal component, the quadrature signal component, the in-phase noise component and the quadrature noise component, respectively. 

With the assumption that signal and noise are totally independent random variables, the received signal can be expressed as: 
\begin{equation}
y=y_I+y_Q
\end{equation}
where,
\begin{equation}
y_I=s_I+n_I          
 \end{equation}
\begin{equation}
y_Q=s_Q+n_Q 
\end{equation}

The SNR estimation of the received signal can be expressed:
\begin{equation}
\rho=S/N	
\end{equation}
where $\rho$ is the estimated SNR value , S is the signal power and N is the noise power.
\subsection {SVR Original Algorithm: }
Signal to Variation Ratio $($SVR$)$ method, as was first explained in ~\cite{8}, is a high order moments based estimation that was originated for channel quality monitoring in multi-path fading channels. It can be applied to channel quality measurement for AWGN channel with an M-PSK modulated signal, however, in general, it is not applicable to other modulation scheme. In~\cite {7}, the authors sketch the derivation for SVR estimation in QPSK modulated signal and assuming AWGN channel. Besides , the authors also showed how the estimation can be applied to real channels. As this paper is concerned with complex channel, this derivation will not be stated. 

SVR estimation is a function of the parameter $\beta$ where $\beta$ is: 
\\
\\

\begin{equation}\label{eq1}
\beta=\frac{E\{ y_n (y_n)^* y_{n-1} y_{n-1}^*\}}  {E\{(y_n y_n^* )^2 \}- E\{y_n(y_n)^* y_{n-1} (y_{n-1})^*\}}
\end{equation}
$y^*$ is the complex conjugate of $y$ and the other two terms, as in~\cite{7}, can be expressed as: 
\begin{equation}\label{eq2}
E\{ y_n (y_n)^* y_{n-1} y_{n-1}^*\} =S^2+2SN+N^2
\end{equation}
\begin{equation}\label{eq3}
\begin{split}
E\{(y_n y_n^* )^2 \}&=E\{|y_n|^4 \}\\&=E\{(y_{In}^2+y_{Qn}^2 )^2\}	
\\&=S^2+4SN+2N^2
\end{split}
\end{equation}
where \eqref{eq2} represent the second moment of the current sample multiplied by the second moment of the previous sample and  \eqref{eq3} is the original representation of the fourth moment in a QPSK modulated signal which is the summation of the in-phase and quadrature components. 

Assuming that signal and noise are independent, substituting \eqref{eq3} and \eqref{eq2} in \eqref{eq1}, and representing $S/N$ as $\rho$, the results can be expressed as: 
\begin{equation}
\beta=\frac{\rho^2 +2\rho+1}  {2 \rho+1}
\end{equation}
One thing to note here is that for modulation schemes other than QPSK, the previous equation will have another representation as the fourth moment ,\eqref{eq3}, will change and the estimation will change accordingly. Finally, solving for $\rho$, the estimation can be expressed as:
\begin{equation}\label{eq5}
\rho= \beta -1+ \sqrt{\beta(\beta-1)}
\end{equation}

In practice, $\beta$ can be calculated by: 
\begin{equation}
\beta= \frac{\frac{1}{(K-1)}\sum_{n=1}^{K-1}|y_n |^2|y_{n-1}|^2}{\frac{1}{(K-1)}\sum_{n=1}^{K-1}|y_n|^4 - \frac{1}{(K-1)} 	\sum_{n=1}^{K-1} |y_n |^2 |y_{n-1}|^2}
\end{equation}
where $K$ is the number of samples. 
\section{Proposed Estimation Algorithm} 
\label{sec:proposed algorithm}
In order to decrease the bias and NMSE of the estimation of the SVR estimation algorithm, let us introduce another form of the fourth moment that considers the difference between in-phase and quadratic components. For a QPSK modulated signal,\cite{3}, this fourth moment can be expressed as: 
\begin{equation}\label{eq4}
\begin{split}
E\{(y_n y_n^* )^2 \}&=E\{|y_n|^4 \}\\&=E\{(y_{In}^2-y_{Qn}^2 )^2\}	
\\&=2SN+N^2
\end{split}
\end{equation}
Substituting \eqref{eq4} and \eqref{eq2} in \eqref{eq1}, considering the signal and noise are independent, and representing $S/N$ as $\rho$, the result can be expressed as:
 \begin{equation}
\beta=\frac{\rho^2 +2\rho+1}  {-\rho^2}
\end{equation}
solving for $\rho$ 
\begin{equation}
\rho= \frac{-1\pm\sqrt{-\beta}}{\beta+1}
\end{equation}

\begin{figure}[ht]
\begin{center}
\includegraphics[width=0.5\textwidth]{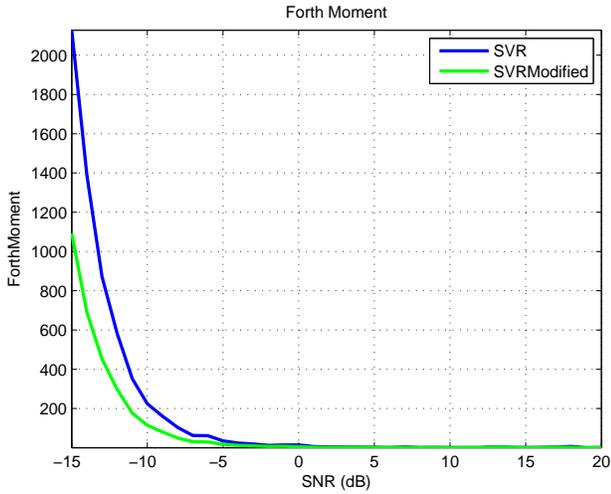}
\end{center}
\caption{Average fourth moment for different SNR in QPSK signal}
\label{F:M4}
\end{figure}

Through simulation, it was noted that the positive root performs good with low SNR values while the negative root had a better estimation with high SNR values. To make use of both roots, a threshold can be chosen to switch between the two roots. The estimation function can now be defined as: 
\begin{equation}
\rho=
\left\{\begin{matrix}
  \frac{-1-\sqrt{-\beta}}{\beta+1}, && SNR>-5
 \\ \frac{-1+\sqrt{-\beta}}{\beta+1},  && SNR \le -5
\end{matrix}\right.
\end{equation}
however, this is impractical in practice as the actual SNR is unknown, so another formulation of the threshold is needed. This can be solved by looking at the fourth moment value instead of the actual SNR value. The fourth moment, as in Fig.~\ref{F:M4}, have the same values in all QPSK signals and hence can be used to set up the threshold in this case. The new estimation algorithm can be expressed as: 
\begin{equation}
\rho=
\left\{\begin{matrix}\label{eq6}
  \frac{-1-\sqrt{-\beta}}{\beta+1}, && |M4|  <20
 \\ \frac{-1+\sqrt{-\beta}}{\beta+1},  && |M4| \ge 20
\end{matrix}\right.
\end{equation}
where $M4$ is the fourth moment that was expressed in \eqref{eq4}.
\section{SIMULATION RESULTS }
\label{sec:results}
Simulation results were done by simulating the new algorithm, equation \eqref{eq6}, SVR original algorithm, equation \eqref{eq5}, and M2M4 algorithm, found in~\cite{9}, which can be expressed as: 
\begin{equation}\label{eq9}
\rho = \frac{\sqrt{2M_2^2-M_4}}{M_2-\sqrt{2M_2^2-M_4}}
\end{equation}
where $M_2$ and $M_4$ are the second and fourth moment that can be expressed in practice, respectively, as:
\begin{equation}
M_2 = \frac{1}{K} \sum_{n=1}^{K}{|y_n|^2}
\end{equation}
\begin{equation}
M_4 = \frac{1}{K} \sum_{n=1}^{K}{|y_n|^4}
\end{equation}
and $K$ is the number of samples. 

Simulations were carried in different scenarios to test algorithms performance with different cases that include testing with low SNR values, high SNR values and low number of samples values. Performance metric used for comparison were the bias of the estimation, or the difference from the true SNR value, and the normalized mean square error $($NMSE$)$ which can be calculated by taking the squared difference between the estimated SNR and the actual one and dividing it by the square value of the actual SNR. In other words: 
\begin{equation}\label{eq7}
NMSE = \frac{\frac{1}{T} \sum(\rho'-\rho)^2}{\rho'^2}
\end{equation}
where $T$ is the number of trials, $\rho'$  is the actual SNR, $\rho$ is the estimated SNR.
 
\subsection {Algorithms Performance at Low SNR Values }
\begin{figure}[ht]
\centering
\includegraphics[width=0.5\textwidth]{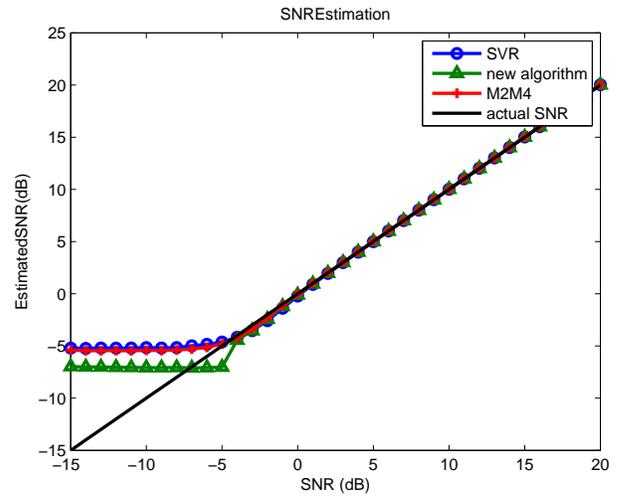}
\caption{Mean for several SNR estimators for $L = 1024$}
\label{F:SNR1024}
\end{figure}

Actual SNR values that were used in simulation ranged from -15 to 20 dB and the simulation was running for 10000 trials to estimate the SNR by the three algorithms. The number of samples was chosen to be 1024 samples. Fig.~\ref{F:SNR1024} shows the mean of the estimated SNR for each algorithm as a function of the SNR where the mean SNR was calculated by averaging 10000 estimates for each algorithm. As shown in the Fig.~\ref{F:SNR1024}, the solid blue line is the ideal SNR estimation that is used here only for comparison. The new algorithm had a better bias estimation compared to others, especially when SNR$<$-5. When SNR$>$-3, all the estimation techniques approached the optimal bias and overlapped with the ideal estimation.
\begin{figure}[ht]
\centering
\includegraphics[width=0.5\textwidth]{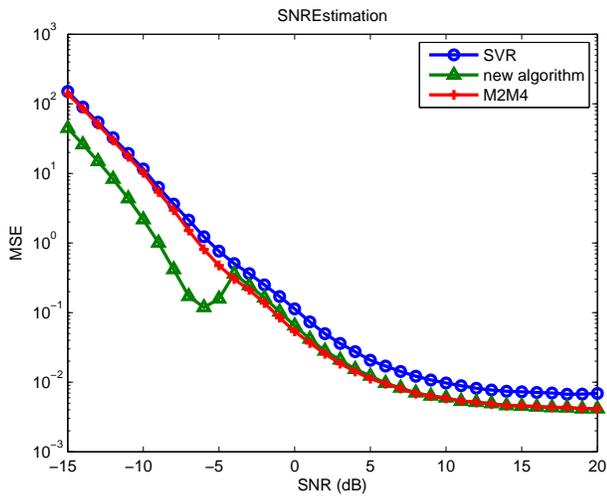}
\caption{NMSE for several SNR values at $L=1024$}
\label{F:MSE1024}
\end{figure}

For the NMSE, as can be seen in Fig.~\ref{F:MSE1024}, the new algorithm had a smaller value in comparison to others when SNR$<$-5 dB. As the SNR value increases, the new algorithm and M2M4 had a similar performance that approached the optimal as their NMSE became too small in the order of 10$^{-2}$. For all SNR values of interest the new algorithm was better than the original SVR algorithm. This can prove the efficiency of the new algorithm compared to others, especially with low SNR values which is the target of this paper.   
\subsection {Algorithm Performance at High SNR Values }
\begin{figure}[ht]
\centering
\includegraphics[width=0.5\textwidth]{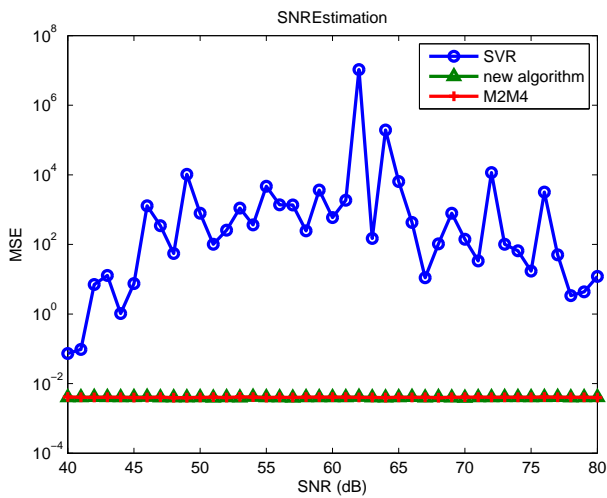}
\caption{NMSE for several SNR at high SNR value for $L=1024$}
\label{F:HMSE}
\end{figure}

Another advantage of the new estimation algorithm was the low NMSE at extremely high SNR values, SNR $>$ 50 dB. At that range, as shown in Fig.~\ref{F:HMSE}, the SVR algorithm performed badly and had a high NMSE while the new algorithm performed almost the same  as M2M4 with relatively good NMSE. 

\begin{figure}[ht]
\centering
\includegraphics[width=0.5\textwidth]{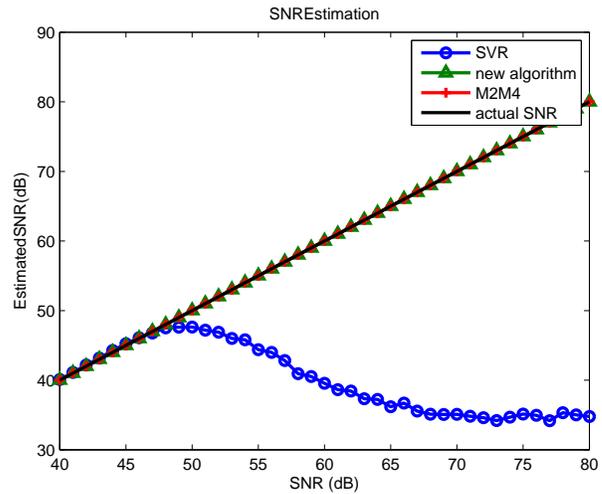}
\caption{Mean for several SNR estimators at high SNR for $L= 1024$}
\label{F:HSNR}
\end{figure}

The reason why SVR algorithm had a bad performance in this case is the mathematical restriction of the algorithm to 50 dB. As shown in Fig.~\ref{F:HSNR}, the algorithm estimated SNR values up to 50 dB and the estimate started decreasing after that. Hence, the difference between the estimated SNR and the actual one increased which resulted in high bias and NMSE increase as well. M2M4 and the new algorithm were not restricted with high SNR value and hence they won’t have a high NMSE in this case. 
\subsection {Algorithms Performance with Different Number of Samples }
\begin{figure}[ht]
\centering
\includegraphics[width=0.5\textwidth]{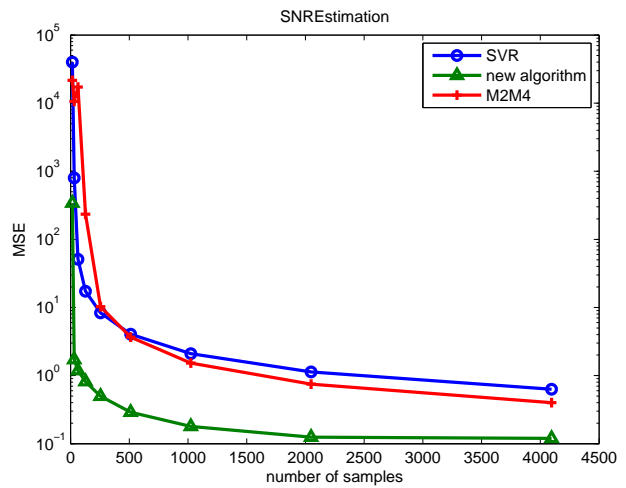}
\caption{NMSE for several number of samples at SNR=-7}
\label{F:MSE-7}
\end{figure}

An important factor in any algorithm design is the number of samples needed by the algorithm to estimate the value correctly. For practical implementation of algorithms in any communication system, it is important for the number of samples to be as small as possible. Hence, the presented three algorithms were tested with different number of samples to check their performance  especially at low number of samples. Fig.~\ref{F:MSE-7} shows the NMSE of the three algorithms at different number of samples with actual SNR value equal to -7. As can be seen, the new algorithm outperformed both M2M4 and SVR especially at low SNR values. 

\begin{figure}[ht]
\centering
\includegraphics[width=0.5\textwidth]{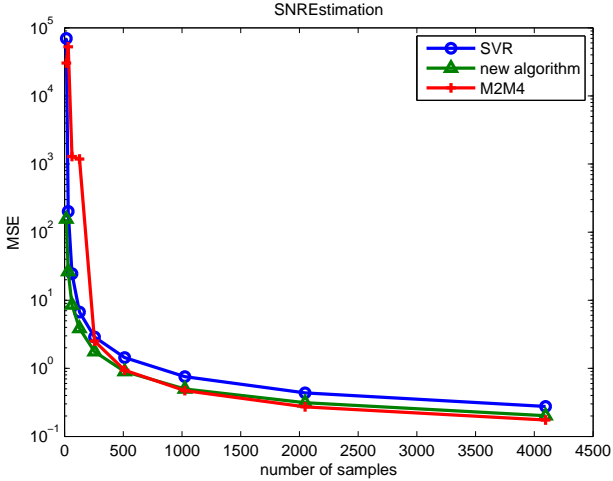}
\caption{NMSE for several number of samples at SNR=-5}
\label{F:MSE-5}
\end{figure}

One thing to note here is that when SNR was -7, M2M4 and SVR estimation performed bad even with high number of samples. However, to check the performance at SNR values where M2M4 would perform good with high number of samples, consider SNR value to be -5dB. At that value, M2M4 algorithm would perform as the new algorithm for number of samples more than 1024. However, as can be seen in Fig.~\ref{F:MSE-5}, the new algorithm had a better performance at low number of samples, when it is less than 512. As the number of samples increase for more than 1024, M2M4 had a slightly better performance than the new algorithm which highlight the effect of number of samples on M2M4 algorithm.     

\begin{figure}[ht]
\centering
\includegraphics[width=0.5\textwidth]{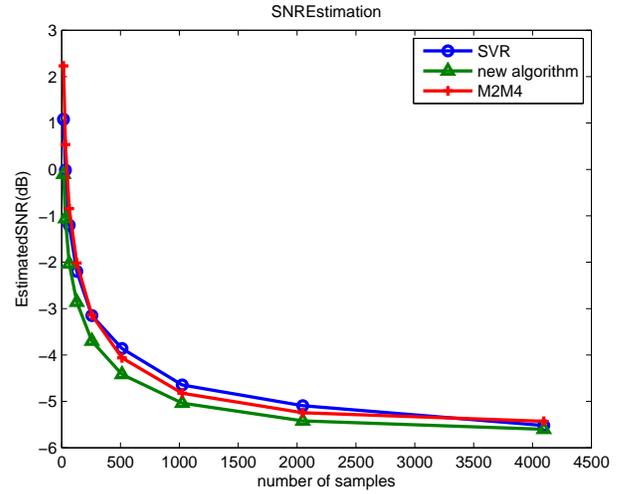}
\caption{SNR estimated value for several number of samples at SNR=-5}
\label{F:SNR-5}
\end{figure}

In such case where the number of samples is less than 512 sample, M2M4 algorithm performed bad as the samples were not enough to estimate the moments value correctly. As a result, M2M4 had wrong estimations that were far away from the actual SNR value, as can be seen from Fig.~\ref{F:SNR-5} where the actual SNR value was -5. Meanwhile, the new algorithm had an estimation that was the closest to the actual SNR value which lead to its NMSE to be the least. However, as the number of samples increased, all the three algorithms came toward -5dB estimation and hence had similar bias and NMSE.   

\begin{figure}[ht]
\centering
\includegraphics[width=0.5\textwidth]{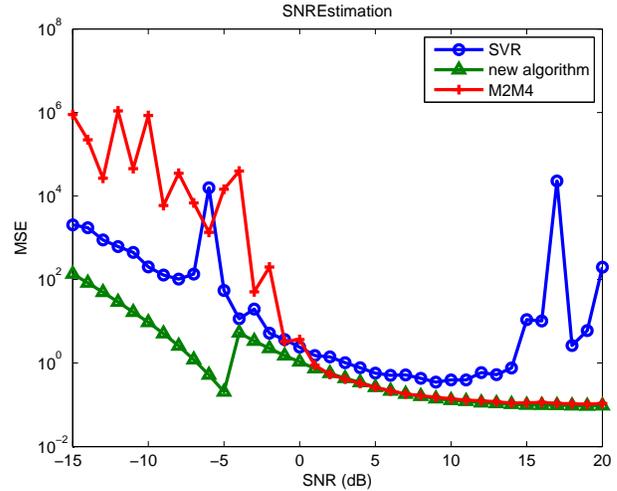}
\caption{NMSE for several SNR at number of samples $L=64$}
\label{F:MSE64}
\end{figure}

To test the performance with different SNR values at low sampling rate, we consider the number of samples to be 64 sample and varied the SNR values from -15 to 15. As can be seen in Fig.~\ref{F:MSE64}, M2M4 algorithm performed even worse than SVR algorithm and kept unstably varying at low SNR. The reason for that performance, as stated before, is that the number of sample are not enough for correct estimation for such low SNR. The new algorithm showed a slightly higher NMSE compared to high number of samples, however the effect was much less than the other two presented algorithms. At higher SNR value, SNR>0, both M2M4 and the new algorithm had the same performance as both estimate where close to the actual SNR value. For SVR original algorithm, low number of samples did not affect the estimation at low SNR values. However, for SNR values more than 10 dB, the NMSE and the bias started to increase and the algorithm performed badly in that case, similar to more than 50 dB performance with high number of samples. 
\subsection{Complexity Added by the Algorithm:} 
The proposed algorithm didn’t add much computational complexity as the operations included within the algorithms were the same except that the new algorithm would compare to a threshold. The computations included multiplications and additions and didn’t need any iteration to get an accurate estimate. Hence, the three presented algorithms have relatively low computational complexity compared to those algorithms that require iterative procedure, such as Maximum Like-hood (ML) [2]. The operational complexity of the three methods is L times, where L is the number of samples. As a result, the overall complexity of the three algorithm would be simply proportional to L and can be simulated or implemented on hardware easily.
\section{Conclusion}
\label{sec:Conclusion}
In this paper, a new SNR estimation technique was proposed on a modification of SVR traditional estimation algorithm. The algorithm formulated another representation of the fourth moment using the difference between the in-phase and quadrature component instead of the summation. This resulted in a better bias and NMSE of the estimates compared to original SVR. Results showed that the new algorithm performs better compared to SVR in all SNR values of interest while it is better than M2M4 at very low SNR values. At a low number of samples, the new algorithm proved its efficiency compared to both M2M4 and SVR, especially with the inaccurate estimation of M2M4 in this case. The results were applied to QPSK modulated signal only, but can be applied to M-PSK with slight modifications in $\beta$ parameter representation. Future enhancements of the algorithm can be by applying it to modulation schemes other than QPSK and see the effect on the estimator and its complexity. Another future enhancement can be by applying the algorithm to Rayleigh faded channels which would include another formulation of the signal that consider the channel effect instead of white Gaussian channel. 
\section*{Acknowledgment}
This publication was made possible by the support of the NPRP grant 5-559-2-227 
from the Qatar National Research Fund(QNRF).  The statements made herein are solely the
responsibility of the authors.
\bibliographystyle{IEEEtran}
\bibliography{paper}
\end{document}